# Improving fretting corrosion resistance of CoCrMo alloy with TiSiN and ZrN coatings for orthopedic applications

Chen-En Tsai[1,6], James Hung[2], Youxin Hu[3], Da-Yung Wang[4], Robert M. Pilliar[3], Rizhi Wang[1,5,6]*

[1]Department of Materials Engineering, University of British Columbia, Vancouver, BC, Canada
[2]Aurora Scientific Corp., Richmond BC, Canada
[3]Institute of Biomaterials and Biomedical Engineering, University of Toronto, Toronto, ON, Canada
[4]Mingdao University, Taiwan, ROC
[5]School of Biomedical Engineering, University of British Columbia, Vancouver, BC, Canada
[6]Centre for Hip Health and Mobility, Vancouver, BC, Canada

* Corresponding address:
Dr. Rizhi Wang
Department of Materials Engineering
University of British Columbia
309-6350 Stores Road
Vancouver, BC V6T 1Z4
Canada
Tel: +1-604-822-9752
Fax: +1-604-822-3619
Email: rzwang@mail.ubc.ca

**Abstract**

Total hip replacement is the most effective treatment for late stage osteoarthritis. However, adverse local tissue reactions (ALTRs) have been observed in patients with modular total hip implants. Although the detailed mechanisms of ALTRs are still unknown, fretting corrosion and the associated metal ion release from the CoCrMo femoral head at the modular junction has been reported to be a major factor. The purpose of this study is to increase the fretting corrosion resistance of the CoCrMo alloy and the associated metal ion release by applying hard coatings to the surface. Cathodic arc evaporation technique (arc-PVD) was used to deposit TiSiN and ZrN hard coatings on to CoCrMo substrates. The morphology, chemical composition, crystal structures and residual stress of the coatings were characterized by scanning electron microscopy, energy dispersive x-ray spectroscopy, and X-ray diffractometry. Hardness, elastic modulus, and adhesion of the coatings were measured by nano-indentation, nano-scratch test, and the Rockwell C test. Fretting corrosion resistance tests of coated and uncoated CoCrMo discs against Ti6Al4V spheres were conducted on a four-station fretting testing machine in simulated body fluid at 1Hz for 1 million cycles. Post-fretting samples were analyzed for morphological changes, volume loss and metal ion release. Our analyses showed better surface finish and lower residual stress for ZrN coating, but higher hardness and better scratch resistance for TiSiN coating. Fretting results demonstrated substantial improvement in fretting corrosion resistance of CoCrMo with both coatings. ZrN and TiSiN decreased fretting volume loss by more than 10 times and 1000 times, respectively. Both coatings showed close to 90% decrease of Co ion release during fretting corrosion tests. Our results suggest that hard coating deposition on CoCrMo alloy can significantly improve its fretting corrosion resistance and could thus potentially alleviate ALTRs in metal hip implants.

# 1. Introduction

Osteoarthritis is one of the most prevailing chronic diseases in the current world. It is a joint disease affecting more than 250 million people worldwide and the number is expected to increase over the next few decades [1]. The most effective treatment for late stage osteoarthritis would be total hip replacement. It is reported that more than 1 million total hip replacements are performed annually worldwide [2]. Current artificial hip implants typically consist of a titanium femoral stem press-fitted into a CoCrMo hemispherical femoral head that articulates with a titanium acetabular cup with a highly crosslinked polyethylene liner at the bearing surface, which is referred to as metal-on-polyethylene (MoP) implants [3]. Although hip arthroplasty is generally a successful procedure, a significant number of patients have been observed with adverse local tissue reactions (ALTRs) to the implants, which are inflammatory lesions that destroy the soft tissues of the hip joint, affecting the prognosis of further clinical solutions [4-7]. Since MoP is the most commonly used hip implant system today [8,9], there is an urgent need to address this issue.

Although the exact mechanism of ALTRs is still unclear, it has been widely attributed to the metal species released from the CoCrMo implant as a result of wear and corrosion at the modular junction where the trunnion of the Ti stem is fitted into the bore of the CoCrMo head [10-13]. This type of wear and corrosion at the head-neck interface has been given the specific term "trunnionosis" [14-16]. Many studies have identified fretting corrosion to be the main degradation mechanism and concern at the modular junction [17-20]. Fretting corrosion is a type of corrosion induced by cyclic micro-abrasion between two contacting surfaces. This would disrupt the formation of passive oxide films on implant surfaces thereby leading to accelerated release of metal ions and metal oxide particles [19,21]. The reformation of oxide layers would result in local oxygen depletion and form a more acidic environment which could further induce crevice and pitting corrosion [22,23]. Therefore, protection on the modular junction surfaces to reduce fretting corrosion should potentially prevent ALTRs.

Researchers have been studying the application of various protective layers to alleviate wear and corrosion on orthopedic implants in order to increase their lifespan since the 1990s [24,25]. Nitride coatings including silicon nitrides [26-28] and superlattice coatings, which are coatings with multiple alternating nitride layers in nanometer scale [29-31], are studied most extensively for biomedical applications. An effective coating must be uniform, corrosion resistant and bond strongly to its substrate. Most importantly, for biomedical application, the coatings need to be biocompatible. Transition metal nitride coatings have been widely studied because of their excellent biocompatibility, corrosion resistance, high hardness and low friction coefficient [32-38]. Hendry and Pilliar [39] previously utilized nitride coatings to specifically study fretting corrosion prevention on Ti6Al4V alloy. Better fretting corrosion resistance with TiN and ZrN coatings by PVD was demonstrated in their work. However, CoCrMo alloy rather than Ti6Al4V has been linked to development of ALTRs in recent clinical studies [7,10,40]. Although several studies have been undertaken on fretting corrosion behavior of CoCrMo alloys [21,41,42], limited studies have yet to explore the strategy of depositing hard coatings on CoCrMo. Therefore, a comprehensive fretting corrosion analysis of coatings on CoCrMo alloy is needed.

In this study, ZrN and TiSiN coatings were chosen to improve fretting corrosion resistance of CoCrMo alloy. ZrN was selected due to its distinguished chemical and physical properties among different transition metal nitrides (TiN, CrN, HfN, TaN, etc) such as high chemical, thermal stability [43] and low electrical resistivity [44]. It has also been proven to be effective in improving wear and corrosion resistance of magnesium alloys [45] and stainless steels [46]. Hubler et al. even observed greater wear and corrosion resistant enhancement of ZrN compared with TiN on 316L femoral implant alloy [47]. For TiN coatings, the latest development has been the incorporation of other elements such as B, C or Si to further improve mechanical properties. These nanocomposite coatings consist of a nano-crystalline phase TiN embedded in an amorphous B, C or Si nitride matrix. These coatings possess exceptionally high hardness, usually in the range of 40 to 50 GPa [48,49]. TiSiN coatings are of particular interest in biomedical applications due to their high hardness, elastic modulus, low friction coefficient and high wear and corrosion resistance [50-52]. Nanocomposite coatings can be synthesized through various physical vapor deposition (PVD) techniques such as magnetron sputtering, cathodic arc evaporation, electron ion plating. Among these techniques, cathodic arc evaporation (arc-PVD) has received extensive attention due to its high ionization rate and high current density which would be expected to lead to high deposition rate and good adhesion of coatings [53].

This study intends to fill in the gap in the literature by addressing the fretting corrosion issue that was commonly found in hip implants. As previously noted, limited studies have been reported on fretting corrosion testing and mechanistic analyses of CoCrMo with hard coatings. One objective of this study is to simulate fretting corrosion that might occur *in-vivo* at the trunnion-femoral head junction of MoP type implants and investigate the possible improvement in fretting corrosion resistance of CoCrMo alloy with either ZrN or TiSiN coatings. The fretting corrosion tests were conducted using Ti6Al4V against unmodified and arc-PVD modified CoCrMo alloy in simulated body fluid (SBF), followed by comprehensive material and mechanistic analyses. To the authors' knowledge, this is one of the first studies in a laboratory setting to recreate similar fretting corrosion results as found in retrieval studies [10]. More importantly, this study proposes and demonstrates the TiSiN nanocomposite coating on the CoCrMo alloy as an effective way of reducing Co ion release caused by fretting corrosion in modular total hip implants.

## 2. Materials and Methods

The CoCrMo alloy (cast ASTM F75-76, Deloro Stellite) used in the experiment was cut into 1 cm lengths from a 16-mm diameter rod. The cut surfaces of the samples to be tested were then ground with a series of sandpapers, polished with 6 and 1 μm diamond suspensions, and finally polished with a 0.05 μm silica suspension to a mirror finish. ZrN and TiSiN coatings were deposited onto the as-polished CoCrMo alloys via arc-PVD (Aurora STAR4 system, Richmond, Canada). Before the deposition, substrate surfaces to be coated were cleaned with Ar ion bombardment for 30 minutes. The temperature during both coating depositions was controlled at 380~420 °C. This deposition temperature should not affect CoCrMo microstructure and properties. For ZrN deposition, the cathode current was kept at 90 A and the bias voltage was decreased gradually from -150 to -80 V through time as $N_2$ pressure was increased from 2 to 3 Pa. To increase the adhesion of TiSiN coating, an interlayer of TiSiCrN was applied. Deposition parameters of this interlayer can be found in the literature [54]. For TiSiN deposition, the cathode current was kept at 85 A with a bias voltage of -80 V.

Coating Surface morphology and coating chemistry were analyzed using a scanning electron microscope (SEM, FEI Quanta 650) equipped with energy dispersive X-ray spectroscopy (EDS). An optical profilometer (Filmetrics Profilm 3D, KLA) was used to evaluate the root mean square surface roughness (Sq) of the substrates and coatings. Crystal structures of the coatings were characterized with an X-ray diffractometer (XRD, Rigaku MultiFlex) in the 2$\theta$ range of 20° to 90°. The diffractometer was operated at 40 kV (Cu K$\alpha$ radiation) with a scanning speed of 1° per second. The thickness of the coatings was determined by an SEM examination of polished cross-sections.

Residual stress of the coatings was measured with Two-dimensional X-ray diffraction (XRD$^2$) technique using a Bruker D8 discover diffractometer equipped with a Pilatus (2D) detector, following a protocol reported in the literature [55]. Briefly, the measurements were performed in the 2$\theta$ range of the selected peaks and samples were tilted 3 steps in the $\psi$-range with $\phi$-angle at 0°, 120°, 240°. As the samples were tilted at different $\psi$ angles, diffraction peaks (2$\theta$) would be observed shifting due to the d-spacing changes caused by residual stress. Residual stress values can then be calculated through Hooke's Law. Quantitative XRD data analysis was performed using the LEPTOS software (Bruker). The peaks were integrated and fitted with Pearson VII function. A biaxial and shear model was used due to the chosen data collection scheme. The stress values were calculated using equations connecting stress tensor components to diffraction cone distortion.

Hardness and elastic modulus of the substrates and coatings were measured using an MTS Nano Indenter XP with a Berkovich tip and the continuous stiffness measurement (CSM) mode. The depth of penetration was controlled to less than 10% of the coating thickness to minimize substrate effects. The adhesion of the coatings was first evaluated using the Rockwell C indentation test (Buehler Macromet II) under the load of 150 kg. All of the indents were examined using optical microscopy (OM, Nikon Eclipse E600) and SEM. The adhesion strength was then classified from the scale of HF1 to HF6 [56]. Nano-scratch tests were also performed using Nano Indenter XP with a cube corner diamond indenter to quantify the adhesion and further study failure mechanisms of the coatings. During scratch testing, the normal load was held constant (20 μN) for the first 100 μm and ramped to a maximum load of 700 mN over the range of 500 μm at the velocity of 10 μm/s. Failure mechanisms and critical load values corresponding to cohesive ($L_{c1}$) and adhesive ($L_{c2}$) failures [57] were evaluated using SEM images.

Fretting corrosion tests on Ti6Al4V-CoCrMo (coated and uncoated) couples were conducted using a four-station fretting apparatus based on linear reciprocating action similar to a previously used machine [39] (Figure 1). An environmental chamber enclosed each individual sample test station, and the fretting tests were carried out in simulated body fluid (SBF). SBF was prepared according to the literature [58] and kept at pH = 7.40. Ti6Al4V spheres (ASTM F136, Dongguan Tengyue, Guangdong, China) 5 mm in diameter were used to wear against both the uncoated and the arc-PVD coated CoCrMo specimens. Nine samples in total were used in fretting corrosion tests (three on the CoCrMo substrate and three on each of the two coated Co alloy groups). The tests were carried out with a normal load of 2.3 N creating maximum Hertzian contact stress of around 744.8 MPa calculated from the equations in the literature [59]. The frequency was set at 1 Hz with a sinusoidal motion of ±25 μm in amplitude generating a total linear fretting amplitude of 50 μm for 1 million cycles. After the tests, the wear scars were examined with SEM and wear products were collected with acetate cellulose replicating film (Ted Pella) and analyzed with EDS. The film was first softened with acetone and then placed on the surface of the fretted CoCrMo specimens. After acetone volatilization, the film was stripped off the specimen so that the wear particles were transferred onto the film. An optical profilometer (Filmetrics Profilm 3D, KLA) was used to measure the wear track after collection of the wear particles and the wear volume was estimated.

The testing solutions were centrifuged at 4000 G-Force through 2K MWCO (molecular weight cut-off) ultrafiltration membrane (Sartorius, Vivacon 2) to remove debris and particles. The filtrates were then analyzed with inductive coupled plasma mass spectrometry (ICP-MS, Agilent 7700x) to measure concentration of Co and Cr ions, which are the main concerns related to ALTRs. Before the measurement, the machine was calibrated using 5-point calibration with single element standard solutions. One-way ANOVA was used for testing concentration differences among three groups (CoCrMo vs ZrN vs TiSiN). A p-value smaller than 0.05 was considered statistically significant.

## 3. Results

### 3.1 Structural analysis

Figure 2a, 2c are SEM secondary electron images showing surface morphologies for ZrN and TiSiN coatings. The ZrN coating shows a substantially denser surface than the TiSiN coating. This can also be observed in the surface roughness values (Sq) in Table 1 measured using optical profilometer. Both coatings show increased surface roughness compared to uncoated CoCrMo substrates. Figure 3 shows EDS spectra used to determine the surface chemistry of the CoCrMo substrate and the two coatings. ZrN coatings showed strong peaks at their major components Zr and N with similar relative atomic concentrations. The signals of the substrate materials (Co, Cr, Mo) were not observed since the coating thickness is larger than sampling depth of EDS (~1 μm). For TiSiN, strong peaks can be seen at Ti and N with around 7 at% Si. The Cr signal from the interlayer was also recorded. SEM on cross-sections (Figure 2b, 2d) revealed the thickness of the coatings, which are 2.37 μm for ZrN and 1.89 μm for TiSiN.

Table 1. Characteristics of ZrN, TiSiN coatings and the substrate CoCrMo

| Sample | Thickness (μm) | Surface roughness (nm) | Young's modulus (GPa) | Hardness (GPa) | Adhesion | Residual stress (GPa) |
|---|---|---|---|---|---|---|
| CoCrMo | N/A | 3.7 ± 0.3 | 180 ± 9 | 10.6 ± 1.2 | N/A | N/A |
| ZrN | 2.37 | 13.3 ± 2.1 | 409 ± 19 | 29.3 ± 2.0 | HF3 | -5.93 ± 0.10 |
| TiSiN | 1.89 | 40.6 ± 1.8 | 396 ± 29 | 41.6 ± 3.2 | HF3 | -8.00 ± 0.02 |

Figure 4 shows the XRD pattern of ZrN and TiSiN coatings on CoCrMo substrate. CoCrMo alloy usually exhibits two different phases which are FCC (γ) and HCP (ε). Peaks of 2θ at 44°, 51.39°, and 75.26° correspond to the γ phase and peak at 47.26° indicates the ε phase as reported in the literature [60]. It is clearly shown that in these samples, the γ-phase has dominance over the ε-phase. ZrN coatings show cubic (B1 NaCl) structure with major diffraction peaks of (111), (200), (311) and (222). It has been reported that nanocomposite TiSiN coating consists of TiN crystalline phases surrounded by an amorphous $Si_3N_4$ region [61]. As can be seen in the XRD pattern, diffraction peaks of TiSiN coatings coincide with cubic TiN (B1 NaCl) structure diffraction peaks of (200), (311) and (222). The grain size of the coating was calculated to be 13 nm from the Scherrer equation.

## 3.2 Mechanical analysis

Two-dimensional X-ray diffraction (XRD$^2$) was performed to measure residual stress in the coatings. (422) and (111) planes, corresponding to $2\theta$ of 110 degrees and 36 degrees, were chosen for ZrN and TiSiN coating, respectively. The XRD$^2$ images at different angles are shown in Figure 5. The shift of the peaks in different $\psi$ angles is evident, indicating high residual stress in the coatings. The positions of the peaks were fitted with Pearson VII function and the shift to higher $2\theta$ angle with increasing $\psi$ implies a decrease in d-spacing, indicating compressive residual stress. The residual stress values were found to be -5.93 GPa for ZrN and -8.00 GPa for TiSiN per Table 1.

Table 1 summarizes hardness and modulus values measured by nano-indentation. Both coatings showed increased hardness compared with the substrate. ZrN and TiSiN displayed similar elastic moduli while TiSiN had substantially higher hardness than ZrN.

Rockwell C test was conducted as a quick assessment of the adhesion of the coatings. The coatings were indented with a cone-shaped diamond tip under a load of 150 kg causing an area of coating damage. The indentation scars were then evaluated with OM and SEM. The damage was classified according to VDI3198 standard from HF1 to HF6, where HF1 demonstrates great adhesion with mild cracking and HF6 indicates unacceptable adhesion with severe delamination around the indentation scar [56]. The image of ZrN coating (Figure 6a, 6b) shows a number of radial cracks at the boundary of indentation scars as well as a few regions in which delamination has occurred. The adhesion strength of this coating can be classified as HF3 suggesting acceptable adhesion according to VDI3198 standard. The TiSiN indentation scar also shows radial cracks surrounding the indentation scar in Figure 6c, 6d. Moreover, ring cracks and several regions of delamination were observed surrounding the indent. Therefore, adhesion strength of TiSiN should also be classified as HF3.

Nano-scratch testing was used to further understand the failure mechanisms and critical load values for both coatings. The typical scratch track of ZrN is shown in Figure 7. Figure 7b shows the starting point of tensile crack propagation, which is identified as $Lc_1$. As load progresses, large areas of compressive delamination, where the substrate is exposed, start to develop. The start of this type of failure corresponds to $Lc_2$ (Figure 7c). For TiSiN, tensile cracks are also presented at $Lc_1$ (Figure 8b). However, a different failure mode is observed at $Lc_2$. Chipping can be seen in figure 8c. No apparent delamination can be observed even though substrate material is exposed. Table 2 summarizes the failure modes and critical load values of TiSiN and ZrN. Critical loads of TiSiN are larger than ZrN for both $Lc_1$ and $Lc_2$. Two coatings display the same mechanism at crack propagation. Severe large area delamination is observed in ZrN as opposed to chipping in TiSiN coating.

Table 2. Critical loads and failure mechanism of ZrN and TiSiN coatings

| Coating | | Critical Loads (mN) | Mechanism |
|---|---|---|---|
| ZrN | $Lc_1$ | 175.9 ± 8.8 | Tensile cracks |
| | $Lc_2$ | 262.3 ± 40.1 | Compressive delamination |
| TiSiN | $Lc_1$ | 207.8 ± 22.1 | Tensile cracks |
| | $Lc_2$ | 329.9 ± 20.6 | Chipping |

### 3.3 Fretting analysis

Figure 9a shows image of wear damage done on a CoCrMo control sample. The wear length appears to be larger than the wear amplitude used. This is caused by the Hertzian contact stress generating a contact area already larger than the total wear amplitude used in the fretting test. Bare CoCrMo substrate group is subjected to significant wear damage with substantial fretting wear particles surrounding the wear perimeter. The wear particles contain mainly Cr, Ca, P, O with depleted Co (Figure 9b), which is similar to corrosion particles obtained from retrieved implants [10]. As can be seen in Figure 10 and Figure 11, the wear damage is largely reduced by the coatings on CoCrMo alloys. However, in Figure 10, the ZrN coating was worn out during fretting test. The backscattered electron image of the wear scar shows depletion of the coating with similar composition to CoCrMo alloy. Two types of fretting wear particles were observed on ZrN-coated CoCrMo. The first type (Figure 10c) contained mainly Zr and O. These particles were detected surrounding the wear scar, which were generated when the coating was still protecting the CoCrMo surface. Another type of particles (Figure 10d) contained Zr and O incorporated with Cr and P, which are the main elements observed with uncoated CoCrMo wear products. This type of wear particles was observed within the wear area indicating its formation after coating depletion. The wear scar of TiSiN coating in Figure 11, on the other hand, showed an intact coating surface from back-scattered images after fretting. Wear particles (Figure 11c) of mainly Ti, Ca, P, O were found on the coating surface.

Wear scars of Ti6Al4V spheres are shown in Figure 12. Interestingly, Ti6Al4V spheres in the bare CoCrMo group displayed the greatest wear damage. The wear damage is substantially reduced in both coating groups with TiSiN having the smaller wear scar. The wear diameters on Ti6Al4V spheres from each group are quantified and presented in Figure 12d. It is worthwhile to note that scratches were observed surrounding the wear areas in both coating groups but not in bare CoCrMo group. Also, more severe scratches were found in TiSiN coating group compared to ZrN. This might be caused by defects that were originally observed on the coating surfaces (Figure 2).

Figure 13 shows results obtained from optical profilometer. Wear damage is largely reduced in the arc-PVD coated groups. However, ZrN showed significantly larger wear volume and deeper wear depth compared to TiSiN. TiSiN demonstrated intact coating surface with the lowest wear volume and only around a 300-nm deep wear scar. Figure 14 shows concentrations of Co and Cr ion that were released during fretting corrosion tests. Significantly lower Co ion was released in two coating groups. Cr ion, however, showed similar concentrations among all groups.

### 4. Discussion

Fretting corrosion and the associated metal ion release from CoCrMo at the modular junction of metal implants are a major concern. In this study, we proposed to protect CoCrMo surface with ZrN and TiSiN coatings. We conducted comprehensive analyses on both coatings followed by fretting corrosion tests to examine the viability of hard coatings. Our results show limited fretting wear in TiSiN coating group and substantial reduction of Co ion release with coatings.

One aim of this study is to simulate fretting corrosion that would occur *in-vivo*. Therefore, a homemade fretting apparatus re-designed from a previous report [39] was used instead of commonly used pin-on-disk tribometer. This device enabled us to perform tests on four samples simultaneously in a simulated physiological environment with a small sliding amplitude. We also specifically chose simulated body fluid in this study to ensure a similar environment to that experienced *in-vivo*. A maximum Hertzian contact stress of around 745 MPa was applied in fretting tests, which is much higher than that experienced in implants [62]. This higher contact stress, however, was proven to be effective in examining not only the two coatings but also CoCrMo substrate in laboratory settings. With the apparatus, we were able to reproduce similar fretting corrosion results (both ion release and particle formation) that were observed from retrieved implants in a clinical study [10,63]. The contact stress applied in our fretting tests is well below the hardness of unmodified and modified CoCrMo. However, severe plastic deformation was observed. This could be due to the multi-asperity contact between the two surfaces [64,65]. Fretting wear was started by the rough surface contact between CoCrMo and Ti alloy at a nanoscale. In an ideal situation, the whole contact area should be in contact. However, the rough nanoparticles on the surfaces were the ones to initiate contact against their counterpart. The nanoscale contact area gave rise to high local contact stress causing local plastic deformation. When plastic deformation happened, more particles were generated and thus leading to the spread of plastic deformation over time.

Elevation of Co and Cr ion concentrations in the human body is a contemporary concern with total hip replacements containing CoCrMo alloy [66]. Theoretically, assuming all the Co and Cr released from CoCrMo alloys during fretting tests turned into ions, the Co/Cr ion concentration ratio should be approximately 2.36/1. However, in the control group, we found that the Co ion concentration is more than 100 times over the Cr ion concentration. High Co/Cr ratios were also extensively reported in the serum of patients with CoCrMo hip implants [7,10,67,68]. This higher than stoichiometric Co/Cr ratio could be due to low solubility of Cr ions in neutral pH environment [69,70] resulting in low but similar Cr ion concentrations in all groups and Cr rich wear particles. This is confirmed in our EDS analyses of the wear particles. The EDS spectrum on the fretting corrosion particles showed mainly Cr with depleted Co (Cr/Co is 3.35/1 in Figure 9b), which is similar to that reported in retrieved hip implant studies [10,71].

The main focus of this study is to investigate the improvement in fretting corrosion resistance of CoCrMo and the subsequent metal ion release due to coating deposition. From Figure 14, Co ion concentrations in the testing fluid are significantly lower in both coating groups showing effective protection by the coatings. This is further supported by the results (Figs. 9-13) that both coatings significantly improved the fretting resistance of the CoCrMo substrate. Interestingly, the difference in Co release between the two coating groups is not apparent even though ZrN coating was depleted after fretting tests. It is likely that the depletion happened close to the end of fretting tests. The improvement in fretting resistance can be directly linked to coating quality that can be assessed through modulus, hardness, adhesion, residual stress, and surface morphologies such as roughness. Hardness and elastic modulus have been proven to be crucial mechanical properties in providing wear resistance of hard coatings. It has been widely acknowledged that the ability of a hard coating to resist crack initiation and propagation is proportional to H/E or $H^3/E^2$ ratio [72,73]. Using hardness and elastic modulus values obtained in Table 1, H/E and $H^3/E^2$ are 0.072 and 0.150 for ZrN while the ratios are 0.105 and 0.459 for TiSiN. It is apparent that the ratios for TiSiN coatings are substantially higher than ZrN. Therefore, better fretting resistance exhibited by TiSiN is not unexpected. For surface morphology of both coatings processed by arc-PVD, SEM images show that TiSiN has substantially higher surface roughness compared to ZrN (Table 1). That could be partially due to the addition of Si into TiN that affects the grain growth during deposition process resulting in more defects on the surface [74].

Adhesion has been a major issue of coatings in clinical applications. Several studies have observed severe third body wear on retrieved implants due to coating delamination [75-77]. In this study, we used both Rockwell C tests and nano-scratch tests to examine the adhesion of coatings, with conflicting results. Even though superior scratch performance was shown by TiSiN with only chipping and small areas of substrate exposure as opposed to large area delamination of ZrN in scratch tests, Rockwell C tests showed only HF3 level for both coatings. Based on the fact that TiSiN coating has much better fretting resistance in our study, it seems that the nano-scratch test is a better tool to evaluate coating adhesion. Rockwell C indentation test, although commonly used for quick adhesion evaluation, is not sensitive enough to show difference in the two coatings studied here. Nevertheless, the HF3 level of adhesion from Rockwell C tests showed that adhesion strength for TiSiN should be further improved for clinical applications. Lowering residual compressive stress could be an approach to enhance coating adhesion strength [78]. Residual stress of ZrN and TiSiN coatings were measured to be -5.93 and -8.00 GPa, respectively. Such high compressive residual stress retained during coating deposition could be a significant driving force for delamination. Several parameters, such as bias voltage and deposition temperature, could be controlled to lower residual stress. In addition, optimized interlayer design could also reduce compressive stress and increase adhesion [79].

There are a couple of limitations in this study. The first one is the PVD technique used in this study and its difficulty in depositing a uniform coating with good adhesion on implants with constrained geometries. Ultimately, for clinical applications, any coating candidate will need to be deposited on the interior bore of the CoCrMo femoral heads. One emerging deposition technology, High power impulse magnetron sputtering (HiPIMS), provides high ionization with high energy due to the pulsed power entering the magnetron [80,81]. The emitted plasma is highly steerable for better out-of-sight coverage and can thus provide dense and uniform coatings onto curved surfaces. Secondly, the fretting corrosion tests were performed using 1 million cycles which is approximately equal to one year of *in vivo* usage. Hip implants are normally expected to sustain more than 10 years of usage. Delamination and complete wear of the coating could happen if fretting corrosion cycles were extended. Future studies is therefore needed using either higher cycles or frequency to confirm the long term performance of the coatings. In terms of biocompatibility, the biggest concern may still be the release of Co ions, which have been identified by many researchers as the main driving force leading to Adverse Local Tissue Reactions. However, the potential side effects caused by the release of nanosized wear particles over a long period from either ZrN or TiSiN coatings remain to be addressed.

## 5. Conclusion
In this study, ZrN and TiSiN coatings were deposited through arc-PVD on CoCrMo substrates to improve fretting corrosion resistance. TiSiN coating possess higher hardness and better adhesion while ZrN coating has better surface finish and lower residual stress. Fretting corrosion tests against Ti6Al4V spheres in simulated body fluid found that uncoated CoCrMo alloy had severe wear damage comparing to the coating groups. TiSiN coatings remained intact after 1 million cycles of fretting tests while ZrN coatings were depleted. Both coatings demonstrated significantly lower Co ion release during fretting corrosion tests. Present study indicates that hard coatings on CoCrMo alloy could significantly enhance its fretting resistance. Further studies on fretting corrosion enhancement with different coating designs are needed to provide the optimal designs for clinical applications.

## 6. Acknowledgements
This study was joint supported by Canadian Institutes of Health Research and Natural Sciences and Engineering Research Council of Canada under the CIHR-NSERC CHRP program. Residual stress analysis of this work was conducted at the Molecular Analysis Facility, a National Nanotechnology Coordinated Infrastructure site at the University of Washington which is supported in part by the National Science Foundation (grant NNCI-1542101), the University of Washington, the Molecular Engineering & Sciences Institute, and the Clean Energy Institute.

## References
[1] D.J. Hunter, D. Schofield, E. Callander, The individual and socioeconomic impact of osteoarthritis, Nat. Rev. Rheumatol. 10 (2014) 437-441. http://doi.org/10.1038/nrrheum.2014.44.
[2] R. Pivec, A.J. Johnson, S.C. Mears, M.A. Mont, Hip arthroplasty, Lancet 380 (2012) 1768-1777. http://doi.org/10.1016/S0140-6736(12)60607-2.
[3] F. Eltit, Q. Wang, R. Wang, Mechanisms of adverse local tissue reactions to hip implants, Front. Bioeng. Biotechnol. 7 (2019) 147. https://doi.org/10.3389/fbioe.2019.00176.
[4] M.R. Whitehouse, M. Endo, S. Zachara, T.O. Nielsen, N.V. Greidanus, B.A. Masri, et al., Adverse local tissue reactions in metal-on-polyethylene total hip arthroplasty due to trunnion corrosion: the risk of misdiagnosis, Bone Joint J. 97-B (2015) 1024–1030. https://doi.org/10.1302/0301-620X.97B8.34682.
[5] N.E. Picardo, H. Al-Khateeb, R.C. Pollock, Atypical pseudotumour after metal-on-polyethylene total hip arthroplasty causing deep venous thrombosis, Hip Int. 21 (2011) 762–765. https://doi.org/10.5301/HIP.2011.8839.
[6] G.S. Matharu, H.G. Pandit, D.W. Murray, A. Judge. Adverse reactions to metal debris occur with all types of hip replacement not just metal-on-metal hips: a retrospective observational study of 3340 revisions for adverse reactions to metal debris from the National Joint Registry for England, Wales, Northe. BMC Musculoskelet Disord 17 (2016) 495-506. https://doi.org/10.1186/s12891-016-1329-8.
[7] F. Eltit, A. Assiri, D. Garbuz, C. Duncan, B. Masri, N. Greidanus, et al., Adverse reactions to metal on polyethylene implants: Highly destructive lesions related to elevated concentration of cobalt and chromium in synovial fluid, J. Biomed. Mater. Res. A 105 (2017) 1876-1886. https://doi.org/10.1002/jbm.a.36057.
[8] National Joint Registry for England, Wales, Northern Ireland and the Isle of Man. National Joint Registry 15[th] Annual Report 2018. https://www.hqip.org.uk/wp-content/uploads/2018/11/NJR-15th-Annual-Report-2018.pdf, 2018 (accessed 6 June 2019)
[9] Canadian Institute for Health Information. Hip and Knee Replacements in Canada, 2016-2017 Canadian Joint Replacement Registry Annual Report.

https://secure.cihi.ca/free_products/cjrr-annual-report-2018-en.pdf, 2018 (accessed 6 June 2019)

[10] H.J. Cooper, R.M. Urban, R.L. Wixson, R.M. Meneghini, J.J. Jacobs, Adverse local tissue reaction arising from corrosion at the femoral neck-body junction in a dual-taper stem with a cobalt-chromium modular neck, J. Bone Joint Surg. Am. 95 (2013) 865-872. https://doi.org/10.2106/JBJS.L.01042.

[11] A.M. Kop, E. Swarts, Corrosion of a hip stem with a modular neck taper junction: a retrieval study of 16 cases, J. Arthroplasty 24 (2009) 1019–1023. https://doi.org/10.1016/j.arth.2008.09.009.

[12] J.P. Collier, V.A. Surprenant, R.E. Jensen, M.B. Mayor, H.P. Surprenant, Corrosion between the components of modular femoral hip prostheses. J. Bone Joint Surgery Br. 74-B (1992) 511–517. https://doi.org/10.1302/0301-620X.74B4.1624507.

[13] E.B. Mathiesen, J.U. Lindgren, G.G. Blomgren, F.P. Reinholt, Corrosion of modular hip prostheses, J. Bone Joint Surg. Br. 73 (1991) 569–575. https://doi.org/10.1302/0301-620X.73B4.2071637.

[14] P.S. Pastides, M. Dodd, K.M. Sarraf, C.A. Willis-Owen, Trunnionosis: A pain in the neck, World J. Orthop. 18 (2013) 161-166. https://doi.org/10.5312/wjo.v4.i4.161.

[15] J.R. Berstock, M.R. Whitehouse, C.P. Duncan, Trunnion corrosion: What surgeons need to know in 2018, Bone Joint J. 100-B (2018) 44-49. https://doi.org/10.1302/0301-620X.100B1.BJJ-2017-0569.R1.

[16] R.M. Shulman, M.G. Zywiel, R. Gandhi, J.R. Davey, D.C. Salonen, Trunnionosis: the latest culprit in adverse reactions to metal debris following hip arthroplasty, Skeletal Radiol. 44 (2015) 433-440. https://doi.org/10.1007/s00256-014-1978-3.

[17] J.L. Gilbert, C.A. Buckley, J.J. Jacobs, K.C. Bertin, M.R. Zernich, Intergranular corrosion-fatigue failure of cobalt-alloy femoral stems. A failure analysis of two implants. J. Bone Joint Surg. Am. 76 (1994) 110–115. https://doi.org/10.2106/00004623-199401000-00014.

[18] J.L. Gilbert, C.A. Buckley, J.J. Jacobs, In vivo corrosion of modular hip prosthesis components in mixed and similar metal combinations. The effect of crevice, stress, motion, and alloy coupling, J. Biomed. Mater. Res. 27 (1993) 1533-1544. https://doi.org/10.1002/jbm.820271210.

[19] N.J. Hallab, J.J. Jacobs, Orthopedic implant fretting corrosion, Corros. Rev. 21 (2003) 182-214. https://doi.org/10.1515/CORRREV.2003.21.2-3.183.

[20] J. Geringer, B. Forest, P. Combrade, Fretting-corrosion of materials used as orthopaedic implants, Wear 259 (2005) 943-951. https://doi.org/10.1016/j.wear.2004.11.027.

[21] V. Swaminathan, J.L. Gilbert, Fretting corrosion of CoCrMo and Ti6Al4V interfaces, Biomaterials 33 (2012) 5487-5503. https://doi.org/10.1016/j.biomaterials.2012.04.015.

[22] S.A. Brown, C.A. Flemming, J.S. Kawalec, H.E. Placko, C. Vassaux, K. Merritt, et al., Fretting corrosion accelerates crevice corrosion of modular hip tapers, J. Appl. Biomater. 6 (1995) 19–26. https://doi.org/10.1002/jab.770060104.

[23] P. Panigrahi, Y. Liao, M.T. Mathew, A. Fischer, M.A. Wimmer, J.J. Jacobs, et al., Intergranular pitting corrosion of CoCrMo biomedical implant alloy, J. Biomed. Mater. Res. B Appl. Biomater. 102 (2014) 850-859. https://doi.org/10.1002/jbm.b.33067.

[24] C. Sella, J.C. Martin, J. Lecoeur, A. Le Chanu, M.F. Harmand, A. Naji, et al., Biocompatibility and corrosion resistance in biological media of hard ceramic coatings

sputter deposited on metal implants, Mater. Sci. Eng. A 139 (1991) 49-57. https://doi.org/10.1016/0921-5093(91)90595-E.

[25] R. Günzel, S. Mändl, E. Richter, A. Liu, B. Tang, P. Chu, Corrosion protection of titanium by deposition of niobium thin films, Surf. Coat. Technol. 116-119 (1999) 1107-1110. https://doi.org/10.1016/S0257-8972(99)00321-7.

[26] M. Pettersson, S. Tkachenko, S. Schmidt, T. Berlind, S. Jacobson, L. Hultman, et al., Mechanical and tribological behavior of silicon nitride and silicon carbon nitride coatings for total joint replacements, J. Mech. Behav. Biomed. 25 (2013) 41-47. https://doi.org/10.1016/j.jmbbm.2013.05.002.

[27] L.C. Filho, S. Schmidt, C. Goyenola, C. Skjoldebrand, H. Engqvist, H. Hogberg, et al., The Effect of N, C, Cr, and Nb Content on Silicon Nitride Coatings for Joint Applications, Materials 13 (2020) 1896. https://doi.org/10.3390/ma13081896.

[28] C. Skjoldebrand, S. Schmidt, V. Vuong, M. Pettersson, K. Grandfield, H. Hogberg, et al., Influence of Substrate Heating and Nitrogen Flow on the Composition, Morphological and Mechanical Properties of SiNx Coatings Aimed for Joint Replacements, Materials 10 (2017) 173. https://doi.org/10.3390/ma10020173

[29] A.V. Garza-Maldonado, M.A.L. Hernandez-Rodriguez, M. Alvarez-Vera, J.A. Ortega-Saenz, A. Perez-Unzueta, R. Cue-Sampedro, Biotribological study of multi-nano-layers as a coating for total hip prostheses, Wear 376-377 (2017) 243-250. https://doi.org/10.1016/j.wear.2016.12.043.

[30] P.E. Hovsepian, A.P. Ehiasarian, Y. Purandare, A.A. Sugumaran, T. Marriott, I. Khan, Development of superlattice CrN/NbN coatings for joint replacements deposited by high power impulse magnetron sputtering, J. Mater. Sci. Mater. Med. 27 (2016) 147. https://doi.org/10.1007/s10856-016-5751-0.

[31] S. Gallegos-Cantu, M.A.L. Hernandez-Rodriguez, E. Garcia-Sanchez, A. Juarez-Hernandez, J. Hernandez-Sandoval, R. Cue-Sampedro, Tribological study of TiN monolayer and TiN/CrN (multilayer and superlattice) on Co–Cr alloy, Wear 330-331 (2015) 439-447. https://doi.org/10.1016/j.wear.2015.02.010.

[32] V. Braic, M. Balaceanu, M. Braic, A. Vladescu, S. Panseri, A. Russo, Characterization of multi-principal-element (TiZrNbHfTa)N and (TiZrNbHfTa)C coatings for biomedical applications, J. Mech. Behav. Biomed. Mater. 10 (2012) 197-205. https://doi.org/10.1016/j.jmbbm.2012.02.020.

[33] I.M. Pohrelyuk, V.M. Fedirko, O.V. Tkachuk, R.V. Proskurnyak, Corrosion resistance of Ti–6Al–4V alloy with nitride coatings in Ringer's solution, Corros. Sci. 66 (2013) 392-398. https://doi.org/10.1016/j.corsci.2012.10.005.

[34] B. Subramanian, C.V. Muraleedharan, R. Ananthakumar, M. Jayachandran, A comparative study of titanium nitride (TiN), titanium oxy nitride (TiON) and titanium aluminum nitride (TiAlN), as surface coatings for bio implants, Surf. Coat. Technol. 205 (2011) 5014-5020. https://doi.org/10.1016/j.surfcoat.2011.05.004.

[35] B. Lohberger, N. Steundl, D. Glaenzer, B. Rinner, N. Donohue, H.C. Lichtenegger, et al., CoCrMo surface modifications affect biocompatibility, adhesion, and inflammation in human osteoblasts, Scientific Reports 10 (2020) 1682. https://doi.org/10.1038/s41598-020-58742-9.

[36] V.Ragone, E. Canciani, C.A. Biffi, R. D'Ambrosi, R. Sanvito, C. Dellavia, et al., CoCrMo alloys ions release behavior by TiNbN coating: an in vitro study, Biomed. Microdevices 21 (2019) 61. https://doi.org/10.1007/s10544-019-0417-6.

[37] J. Doring, M. Crackau, C. Nestler, F. Welzel, J. Bertrand, C.H. Lohmann, Characteristics of different cathodic arc deposition coatings on CoCrMo for biomedical applications, J. Mech. Behav. Biomed. 97 (2019) 212-221. https://doi.org/10.1016/j.jmbbm.2019.04.026.

[38] M. Dinu, I. Pana, P. Scripca, I.G. Sandu, C. Vitelaru, A. Vladescu, Improvement of CoCr Alloy Characteristics by Ti-Based Carbonitride Coatings Used in Orthopedic Applications, Materials 10 (2020) 495. https://doi.org/10.3390/coatings10050495.

[39] J.A. Hendry, R.M. Pilliar, The fretting corrosion resistance of PVD surface-modified orthopedic implant alloys, J. Biomed. Mater. Res. 58 (2001) 156–166. https://doi.org/10.1002/1097-4636(2001)58:2<156::AID-JBM1002>3.0.CO;2-H.

[40] H.J. Cooper, C.J. Della Valle, R.A. Berger, M. Tetreault, W.G. Paprosky, S.M. Sporer, et al., Corrosion at the head-neck taper as a cause for adverse local tissue reactions after total hip arthroplasty, J. Bone Joint Surg. Am. 94 (2012) 1655-1661. https://doi.org/10.2106/JBJS.K.01352.

[41] E.K. Ocran, L.E. Guenther, J. Brandt, U. Wyss, O.A. Ojo, Corrosion and fretting corrosion studies of medical grade CoCrMo alloy in a clinically relevant simulated body fluid environment, Metall. Mater. Trans. A 46 (2015) 2696-2709. https://doi.org/10.1007/s11661-015-2834-3.

[42] D. Sun, J.A. Wharton, R.J.K. Wood, L. Ma, W.M. Rainforth, Microabrasion–corrosion of cast CoCrMo alloy in simulated body fluids, Tribol. Int. 42 (2009) 99-110. https://doi.org/10.1016/j.triboint.2008.05.005.

[43] D. Wu, Z. Zhang, W. Fu, X. Fan, H. Guo, Structure, electrical and chemical properties of zirconium nitride films deposited by dc reactive magnetron sputtering, Appl. Phys. A 64 (1997) 593-595. https://doi.org/10.1007/s003390050522.

[44] C.C. Wang, S.A. Akbar, W. Chen, V.D. Patton, Electrical properties of high-temperature oxides, borides, carbides, and nitrides, J. Mater. Sci. 30 (1995) 1627-1641. https://doi.org/10.1016/j.wear.2007.03.014.

[45] Y. Xin, C. Liu, K. Huo, G. Tang, X. Tian, P.K. Chu, Corrosion behavior of ZrN/Zr coated biomedical AZ91 magnesium alloy, Surf. Coat. Technol. 203 (2009) 2554–2557. https://doi.org/10.1016/j.surfcoat.2009.02.074.

[46] W.-J. Chou, G.-P. Yu, J.-H. Huang, Corrosion resistance of ZrN films on AISI 304 stainless steel substrate, Surf. Coat. Technol. 167 (2003) 59–67. https://doi.org/10.1016/s0257-8972(02)00882-4.

[47] R. Hübler, A. Cozza, T.L. Marcondes, R.B. Souza, F.F. Fiori, Wear and corrosion protection of 316-L femoral implants by deposition of thin films, Surf. Coat. Technol. 142-144 (2001) 1078–1083. https://doi.org/10.1016/S0257-8972(01)01321-4.

[48] P. Karvankova, M.G.J. Veprek-Heijman, D. Azinovic, S. Veprek, Properties of superhard nc-TiN/a-BN and nc-TiN/a-BN/a-TiB$_2$ nanocomposite coatings prepared by plasma induced chemical vapor deposition, Surf. Coat. Technol. 200 (2006) 2978-2989. https://doi.org/10.1016/j.surfcoat.2005.01.003.

[49] S. Ma, J. Prochazka, P. Karvankova, Q. Ma, X. Niu, X. Wang, et al., Comparative study of the tribological behaviour of superhard nanocomposite coatings nc-TiN/a-Si$_3$N$_4$ with TiN, Surf. Coat. Technol. 194 (2005) 143-148. https://doi.org/10.1016/j.surfcoat.2004.05.007.

[50] M. Diserens, J. Patscheider, F. Levy, Improving the properties of titanium nitride by incorporation of silicon, Surf. Coat. Technol. 108-109 (1998) 241-246. https://doi.org/10.1016/S0257-8972(98)00560-X.

[51] Y.H. Cheng, T. Browne, B. Heckerman, E.I. Meletis, Mechanical and tribological properties of nanocomposite TiSiN coatings, Surf. Coat. Technol. 204 (2010) 2123-2129. https://doi.org/10.1016/j.surfcoat.2009.11.034.

[52] S. Balasubramanian, A. Ramadoss, A. Kobayashi, J. Muthirulandi, Nanocomposite Ti-Si-N coatings deposited by reactive dc magnetron sputtering for biomedical applications, J. Am. Ceram. Soc. 95 (2012) 2746-2752. https://doi.org/10.1111/j.1551-2916.2011.05029.x.

[53] S.-M. Yang, Y.-Y. Chang, D.-Y. Wang, D.-Y. Lin, W.T. Wu, Mechanical properties of nano-structured Ti-Si-N films synthesized by cathodic arc evaporation, J. Alloys Compd. 440 (2007) 375-359. https://doi.org/10.1016/j.jallcom.2006.12.124.

[54] S.-M. Yang, Y.-Y. Chang, D.-Y. Lin, D.-Y. Wang, W. Wu, Mechanical and tribological properties of multilayered TiSiN/CrN coatings synthesized by a cathodic arc deposition process, Surf. Coat. Technol. 202 (2008) 2176-2181. https://doi.org/10.1016/j.surfcoat.2007.09.004.

[55] B.B. He, Two-Dimensional X-Ray Diffraction, John Wiley & Sons, New Jersey, 2009, pp. 249-328.

[56] Verein Deutscher Ingenieure Normen, VDI 3198, VDI-Verlag, Dusseldorf, 1991.

[57] S.T. Gonczy, N. Randall, An ASTM standard for quantitative scratch adhesion testing of thin, hard ceramic coatings, Int. J. Appl. Ceram. Tec. 2 (2005) 422-428. https://doi.org/10.1111/j.1744-7402.2005.02043.x.

[58] T. Kokubo, H. Takadama, How useful is SBF in predicting in vivo bone activity?, Biomaterials 27 (2006) 2907-2915. https://doi.org/10.1016/j.biomaterials.2006.01.017.

[59] R.G. Budynas, J.K. Nisbett, Shigley's mechanical engineering design, McGraw-Hill, New York, 2011. pp.122-126.

[60] N. Moharrami, D.J. Langton, O. Sayginer, S.J. Bull, Why does titanium alloy wear cobalt chrome alloy despite lower bulk hardness: A nanoindentation study?, Thin Solid Films 549 (2013) 79-86. https://doi.org/10.1016/j.tsf.2013.06.020.

[61] C.-L. Chang, C.-T. Lin, P.-C. Tsai, W.-Y. Ho, D.-Y. Wang, Influence of bias voltages on the structure and wear properties of TiSiN coating synthesized by cathodic arc plasma evaporation, Thin Solid Films 516 (2008) 5324-5329. https://doi.org/10.1016/j.tsf.2007.07.087.

[62] R. English, A. Ashkanfar, G. Rothwell, A computational approach to fretting wear prediction at the head-stem taper junction of total hip replacements, Wear 338-339 (2015) 210-220. https://doi.org/10.1016/j.wear.2015.06.016.

[63] Q. Wang, F. Eltit, D. Garbuz, C. Duncan, B.A. Masri, N. Greidanus, et al., CoCrMo metal release in metal-on- highly crosslinked polyethylene hip implants, J. Biomed. Mater. Res. B Appl. Biomater. 108 (2020) 1213-1228. http://doi.org/10.1002/jbm.b.34470.

[64] L. Si, X. Wang, G. Xie, N. Sun, Nano-adhesion and friction of multi-asperity contact: a molecular dynamics simulation study, Surf. Interface Anal. 47 (2015) 919-925. https://doi.org/10.1002/sia.5797.

[65] L. Frerot, R. Aghababaei, J.-F. Molinari, A mechanistic understanding of the wear coefficient: From single to multiple asperities contact, J. Mech. Phys. Solids 114 (2018) 172-184. https://doi.org/10.1016/j.jmps.2018.02.015.

[66] A.W. Schaffer, A. Pilger, C. Engelhardt, K. Zweymueller, H.W. Ruediger, Increased Blood Cobalt and Chromium After Total Hip Replacement, J. Toxicol. Clin. Toxicol. 37 (1999) 839-844. https://doi.org/10.1081/CLT-100102463.

[67] I.P.S. Gill, J. Webb, K. Sloan, R.J. Beaver, Corrosion at the neck-stem junction as a cause of metal ion release and pseudortumour formation, J. Bone Joint Surgery Br. 94-B (2012) 895-900. https://doi.org/10.1302/0301-620X.94B7.29122.

[68] Y.A. Fillingham, C.J. Della Valle, D.D. Bohl, M.P. Kelly, D.J. Hall, R. Pourzal, et al., Serum metal levels for diagnosis of adverse local tissue reactions secondary to corrosion in metal-on-polyethylene total hip arthroplasty, J. Arthroplasty 32 (2017) S272-277. https://doi.org/10.1016/j.arth.2017.04.016.

[69] D. Rai, B.M. Sass, D.A. Moore, Chromium (III) hydrolysis constants and solubility of chromium(III) hydroxide, Inorg. Chem. 26 (1987) 345–349. https://doi.org/10.1021/ic00250a002.

[70] D. Rai, D.A. Moore, N.J. Hess, L. Rao, S.B. Clark, Chromium(III) hydroxide solubility in the aqueous Na-OH-H2PO4-HPO4-PO4-H2O system: A thermodynamic model, J. Solution Chem. 33 (2004) 1213–1242. https://doi.org/10.1007/s10953-004-7137-z.

[71] M. Huber, G. Reinisch, G. Trettenhahn, K. Zweymuller, F. Lintner, Presence of corrosion products and hypersensitivity-associated reactions in periprosthetic tissue after aseptic loosening of total hip replacements with metal bearing surfaces, Acta Biomater. 5 (2009) 172-180. https://doi.org/10.1016/j.actbio.2008.07.032.

[72] A. Leyland, A. Matthews, On the significance of the H/E ratio in wear control: a nanocomposite coating approach to optimised tribological behaviour, Wear 246 (2000) 1-11. https://doi.org/10.1016/S0043-1648(00)00488-9.

[73] J. Musil, F. Kunc, H. Zeman, H. Polakova, Relationships between hardness, Young's modulus and elastic recovery in hard nanocomposite coatings, Surf. Coat. Technol. 154 (2002) 304-313. https://doi.org/10.1016/S0257-8972(01)01714-5.

[74] C.T. Guo, D. Lee, P.C. Chen, Deposition of TiSiN coatings by arc ion plating process, Appl. Surf. Sci. 254 (2008) 3130-3136. https://doi.org/10.1016/j.apsusc.2007.10.079.

[75] L. Lapaj, J. Wendland, J. Markuszewski, A. Mroz, T. Wisniewski, Retrieval analysis of titanium nitride (TiN) coated prosthetic femoral heads articulating with polyethylene, J. Mech. Behav. Biomed. Mater. 55 (2015) 127-139. https://doi.org/10.1016/j.jmbbm.2015.10.012.

[76] M.K. Harman, S.A. Banks, W.A. Hodge, Wear analysis of a retrieved hip implant with titanium nitride coating, J. Arthroplasty 12 (1997) 938-945. https://doi.org/10.1016/S0883-5403(97)90164-9.

[77] G. Taeger, L.E. Podleska, B. Schmidt, M. Ziegler, D. Nast-Kolb, Comparison of diamond-like-carbon and alumina-oxide articulating with polyethylene in total hip arthroplasty, Mater. Sci. Eng. Technol. 34 (2003) 1094-1100. https://doi.org/10.1002/mawe.200300717.

[78] R. Ali, M. Sebastiani, E. Bemporad, Influence of Ti–TiN multilayer PVD-coatings design on residual stresses and adhesion, Mater. Des. 75 (2015) 47-56. https://doi.org/10.1016/j.matdes.2015.03.007.

[79] C. Mendibide, P. Steyer. J. Fontaine, P. Goudeau, Improvement of the tribological behaviour of PVD nanostratified TiN/CrN coatings — An explanation, Surf. Coat. Technol. 201 (2006) 4119-4124. https://doi.org/10.1016/j.surfcoat.2006.08.013.

[80] K. Sarakinos, J. Alami, S. Konstantinidis, High power pulsed magnetron sputtering: A review on scientific and engineering state of the art, Surf. Coat. Technol. 204 (2010) 1661-1684. https://doi.org/10.1016/j.surfcoat.2009.11.013.

[81] D. Lundin, K. Sarakinos, An introduction to thin film processing using high-power impulse magnetron sputtering, J. Mater. Res. 27 (2012) 780-792. https://doi.org/10.1557/jmr.2012.8.

**Figure legends**

Figure 1. (a) Side view of four-station fretting apparatus. (b) Illustration of the setup of each station.

Figure 2. SEM surface morphology of (a) ZrN and (c) TiSiN coatings. Cross-sectional backscattered electron (BSE) images of (b) ZrN and (d) TiSiN coatings.

Figure 3. EDS spectra of (a) CoCrMo substrate, (b) ZrN and (c) TiSiN coatings deposited on CoCrMo.

Figure 4. XRD spectra of (a) ZrN and (b) TiSiN coatings deposited on CoCrMo.

Figure 5. $XRD^2$ images of (a) ZrN in the gamma range of -76°~ -102° and (b) TiSiN in the gamma range of -66°~ -91° at different $\psi$ angles.

Figure 6. OM images of Rockwell indents of (a) ZrN and (c) TiSiN. SEM images of cracks and delamination spots surrounding Rockwell indents of (b) ZrN and (d) TiSiN.

Figure 7. SEM images of (a) a typical nano-scratch track of ZrN with (b) tensile cracks at $Lc_1$ and (c) compressive delamination at $Lc_2$.

Figure 8. SEM images of (a) a typical nano-scratch track of TiSiN with (b) tensile cracks at $Lc_1$ and (c) chipping and buckling at $Lc_2$.

Figure 9. (a) SEM image of CoCrMo sample after fretting test. (b) EDS spectrum of the fretting corrosion products.

Figure 10. (a) Secondary electron (SE) and (b) Backscattered electron (BSE) images of ZrN coating on CoCrMo after fretting test. EDS spectra of fretting corrosion products formed (c) before (triangle spot) and (d) after (star spot) depletion of the coating.

Figure 11. (a) SE and (b) BSE images of TiSiN coating after fretting test. (c) EDS spectrum of the fretting corrosion products.

Figure 12. SEM images of Ti6Al4V spheres after fretting against (a) CoCrMo, (b) ZrN and (c) TiSiN. (d) Comparison of wear diameter on Ti6Al4V spheres from each group.

Figure 13. (a) Wear volume and (b) Wear track comparison of CoCr, ZrN and TiSiN after fretting against Ti6Al4V.

Figure 14. Concentrations of Co and Cr ion released during fretting test. Please note the different scales used for plotting Co and Cr ions. $p < 0.05$ is marked by *.

Figure

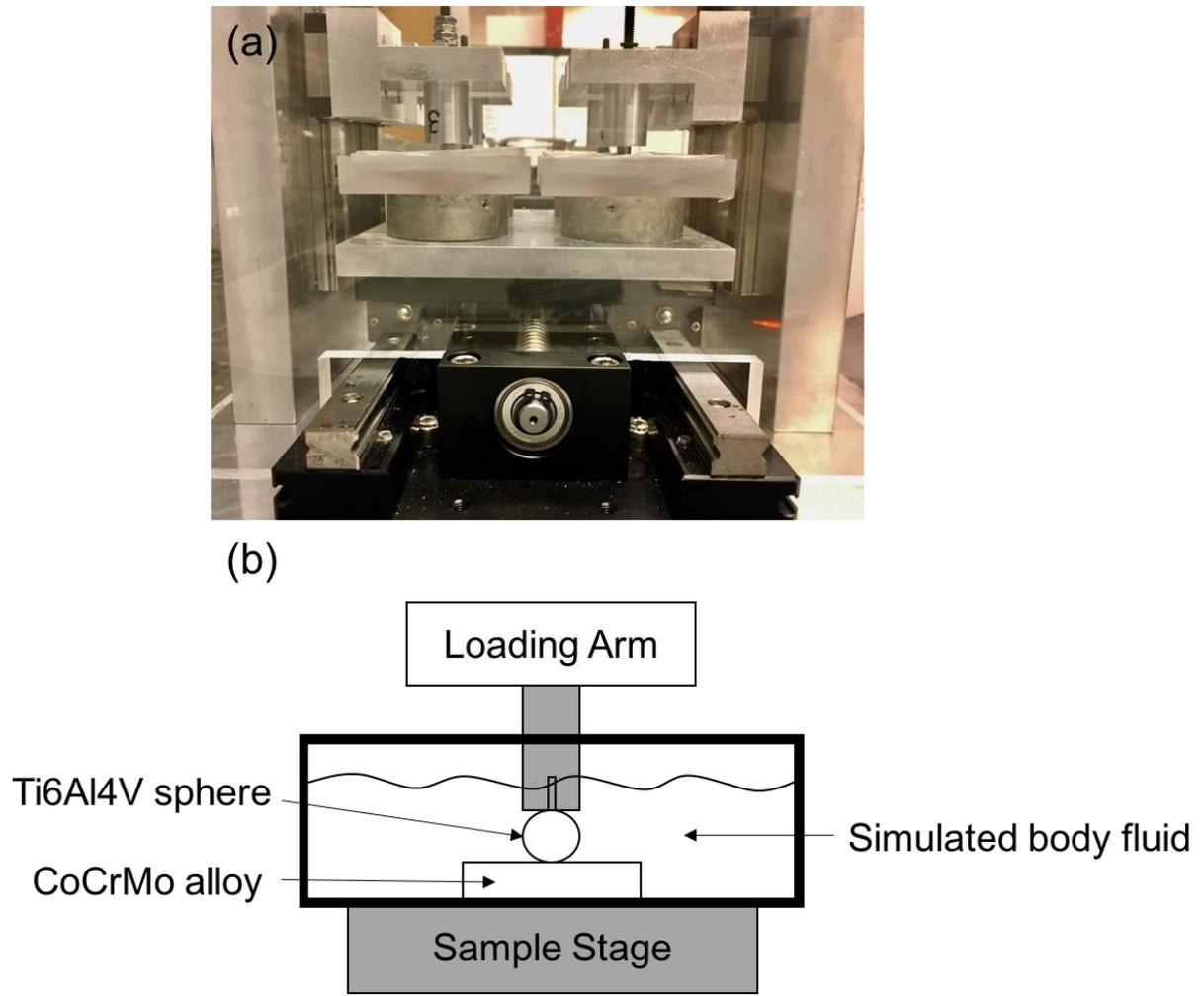

Figure 1. (a) Side view of four-station fretting apparatus. (b) Illustration of the setup of each station.

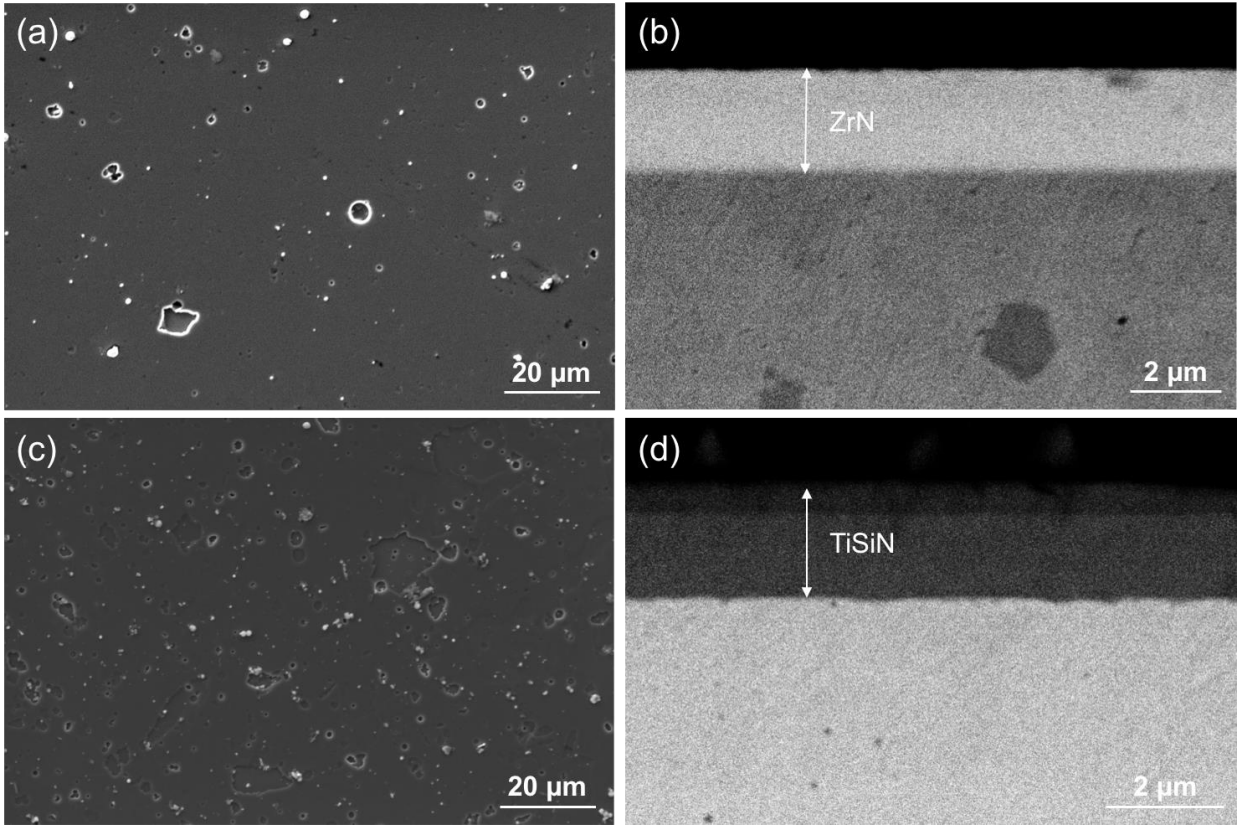

Figure 2. SEM surface morphology of (a) ZrN and (c) TiSiN coatings.  Cross-sectional backscattered electron (BSE) images of (b) ZrN and (d) TiSiN coatings.

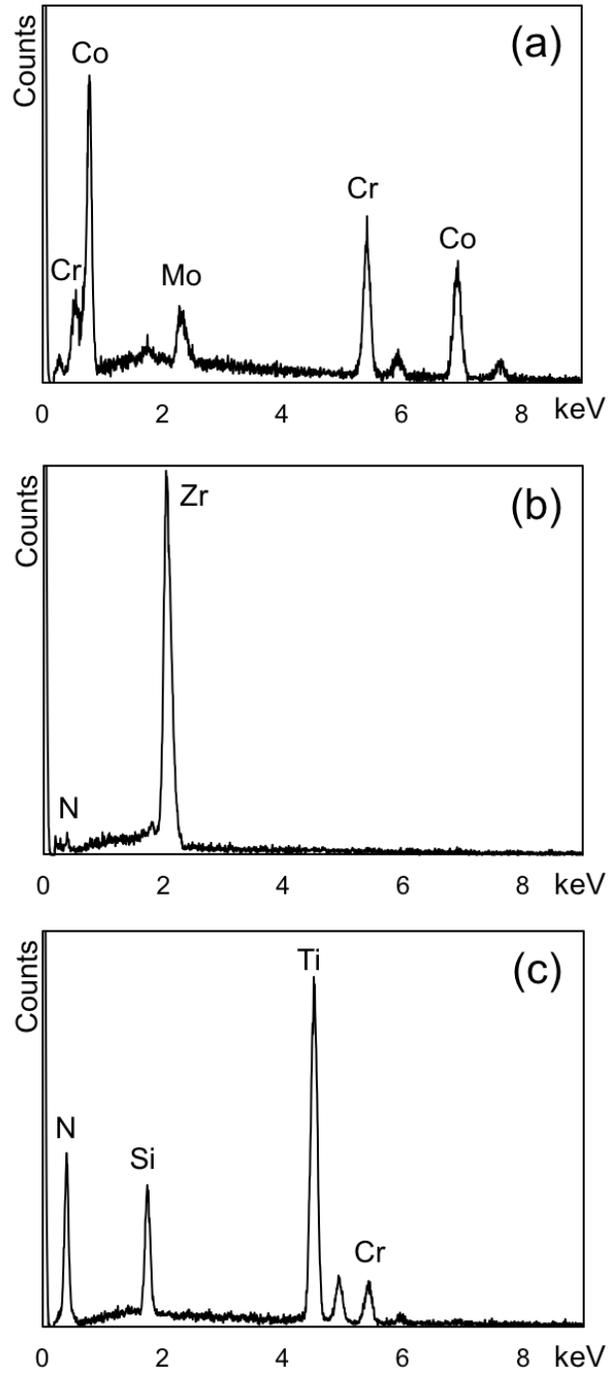

Figure 3. EDS spectra of (a) CoCrMo substrate, (b) ZrN and (c) TiSiN coatings deposited on CoCrMo.

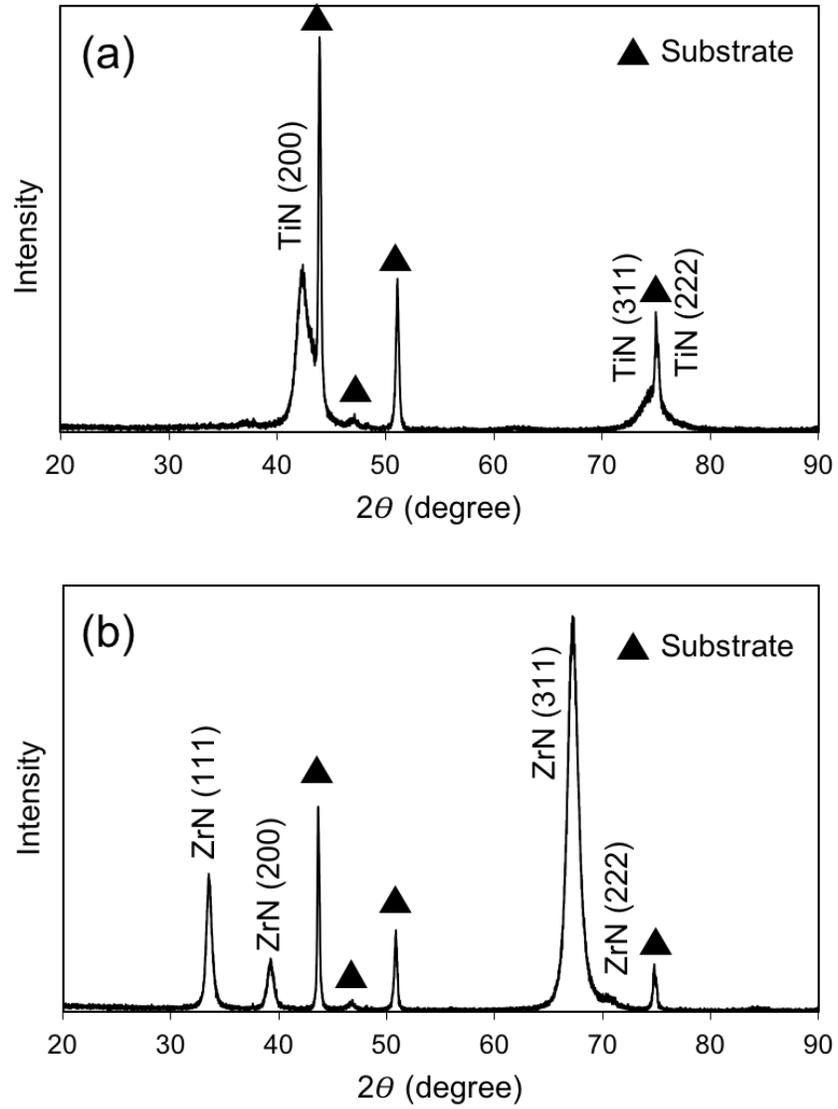

Figure 4. XRD spectra of (a) ZrN and (b) TiSiN coatings deposited on CoCrMo. Substrate peaks are labelled with triangles.

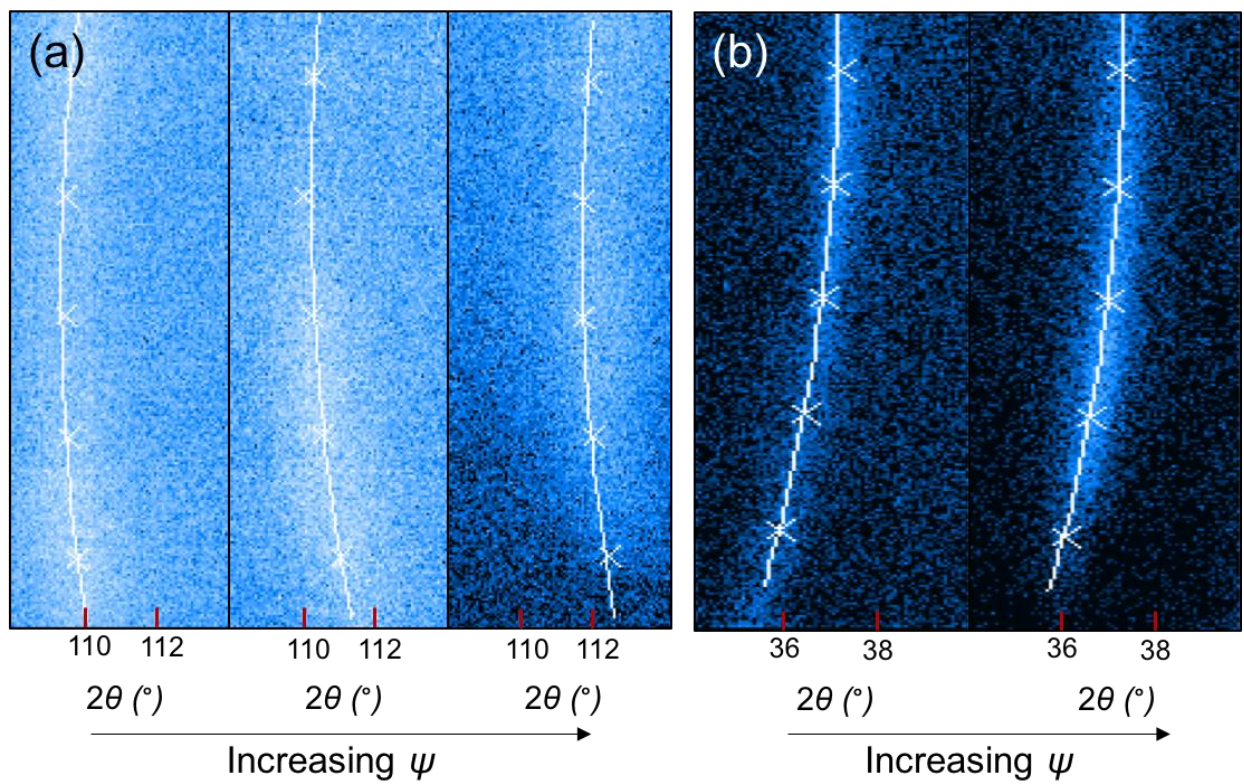

Figure 5. XRD² images of (a) ZrN in the gamma range of -76°~ -102° and (b) TiSiN in the gamma range of -66°~ -91° at different ψ *angles*.

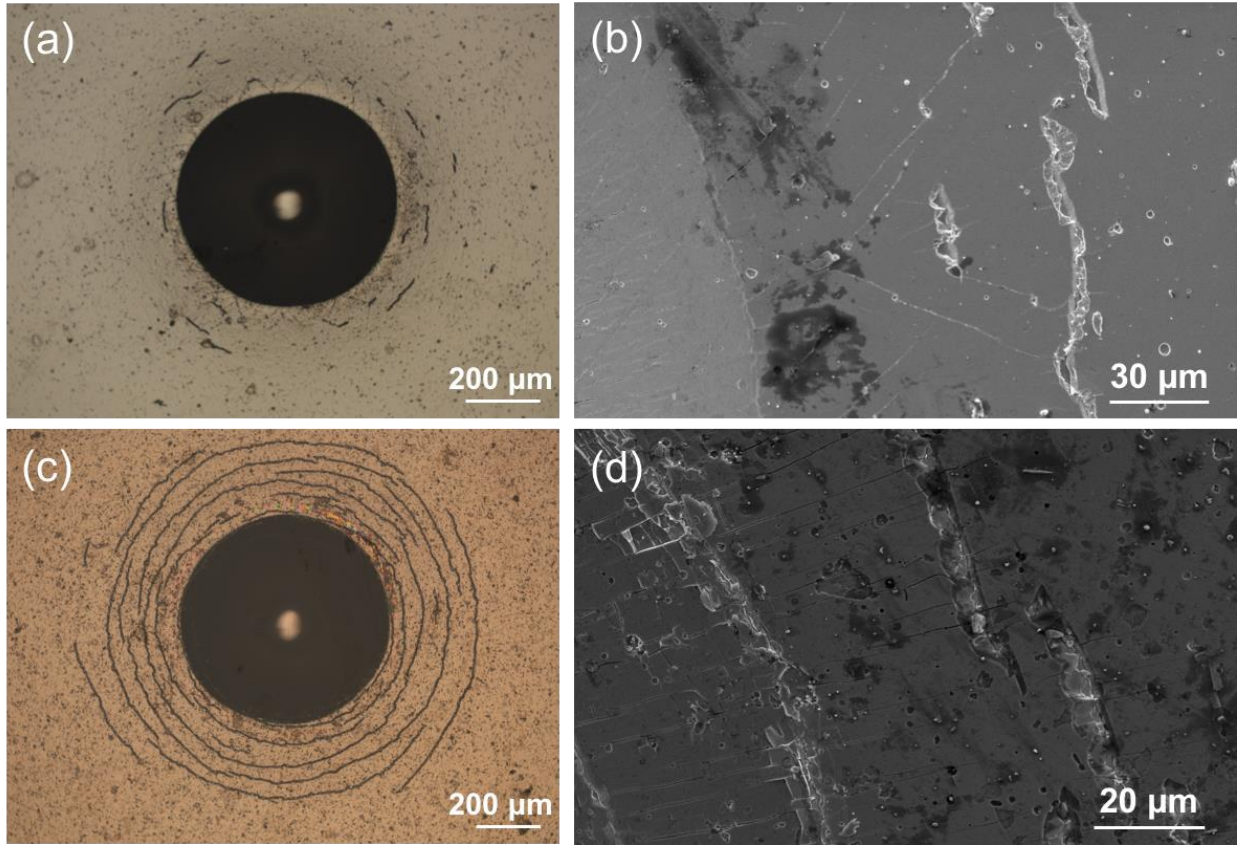

Figure 6. OM images of Rockwell indents of (a) ZrN and (c) TiSiN. SEM images of cracks and delamination spots surrounding Rockwell indents of (b) ZrN and (d) TiSiN.

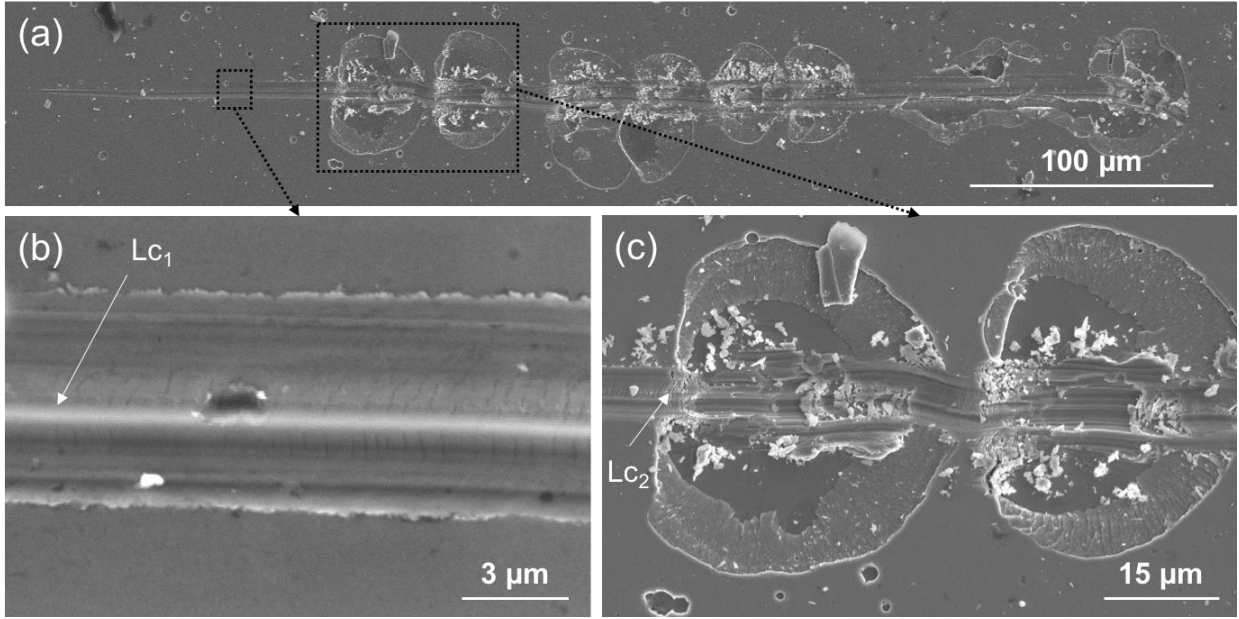

Figure 7. SEM images of (a) a typical nano-scratch track of ZrN with (b) tensile cracks at $Lc_1$ and (c) compressive delamination at $Lc_2$.

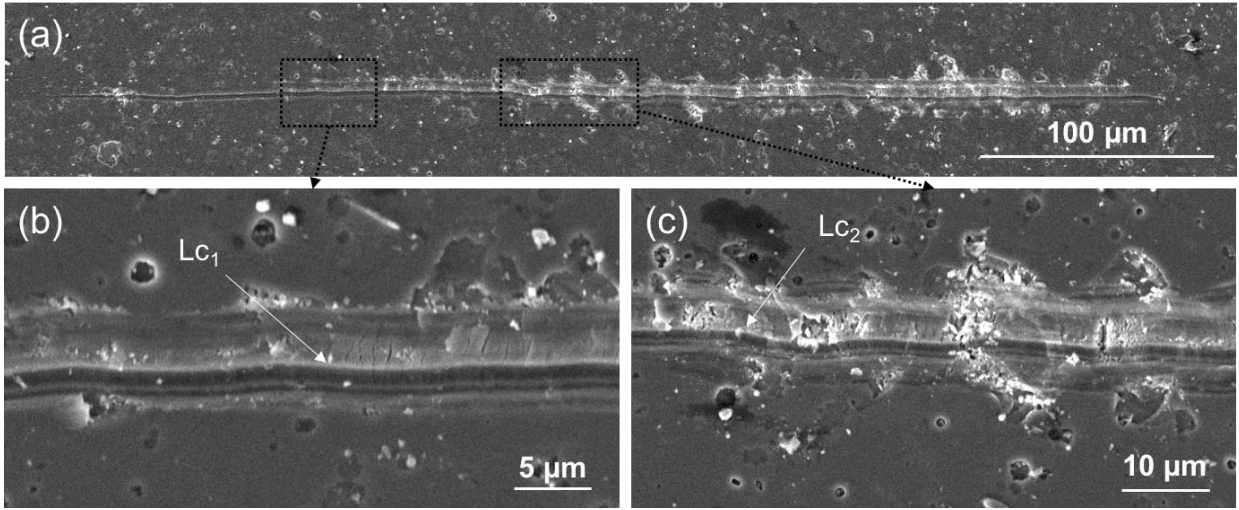

Figure 8. SEM images of (a) a typical nano-scratch track of TiSiN with (b) tensile cracks at $Lc_1$ and (c) chipping and buckling at $Lc_2$.

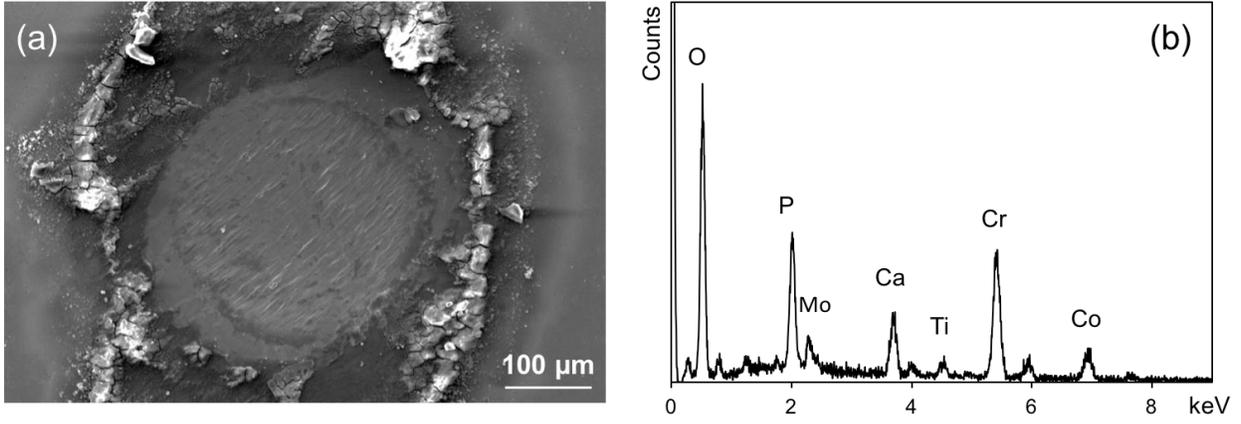

Figure 9. (a) SEM image of CoCrMo sample after fretting test. (b) EDS spectrum of the fretting corrosion products.

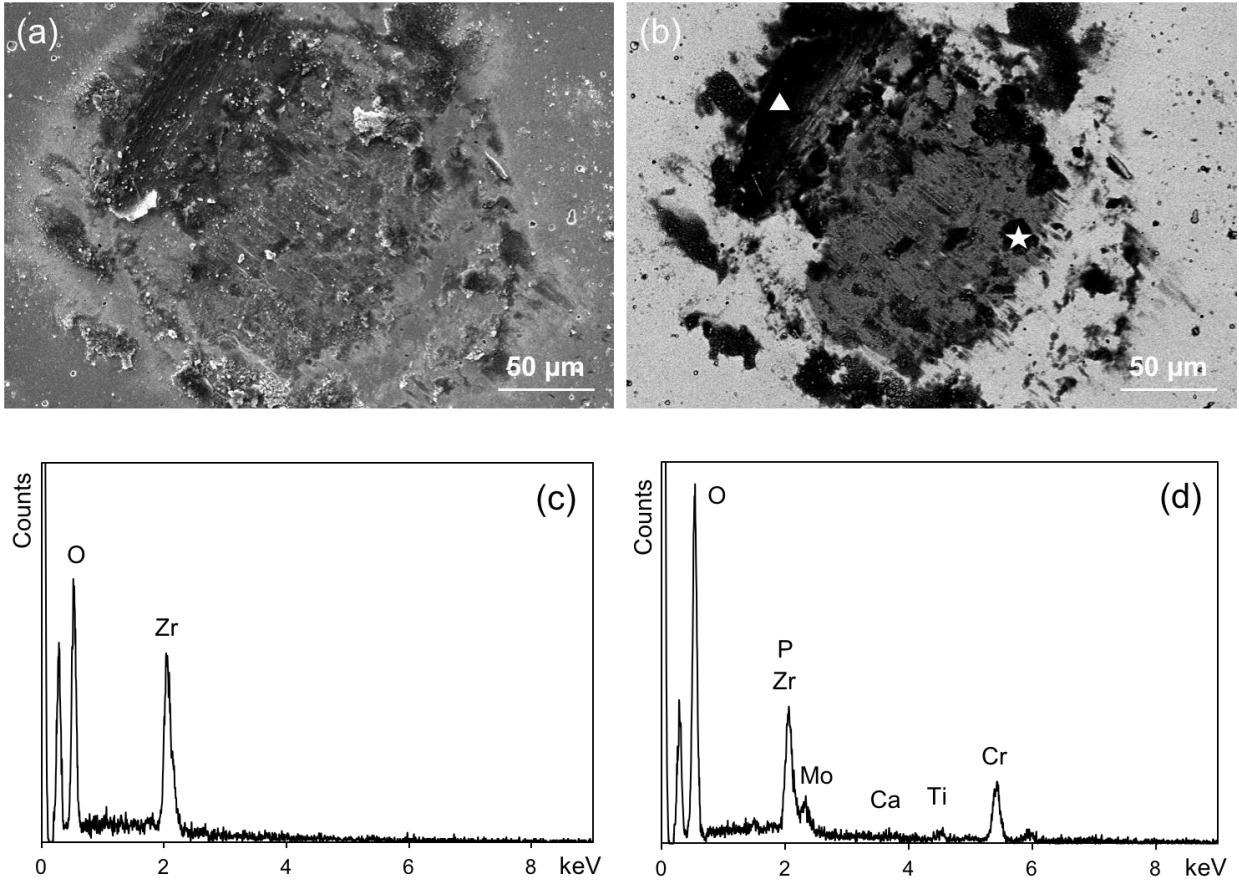

Figure 10. (a) Secondary electron (SE) and (b) Backscattered electron (BSE) images of ZrN coating on CoCrMo after fretting test. EDS spectra of fretting corrosion products formed (c) before (triangle spot) and (d) after (star spot) depletion of the coating.

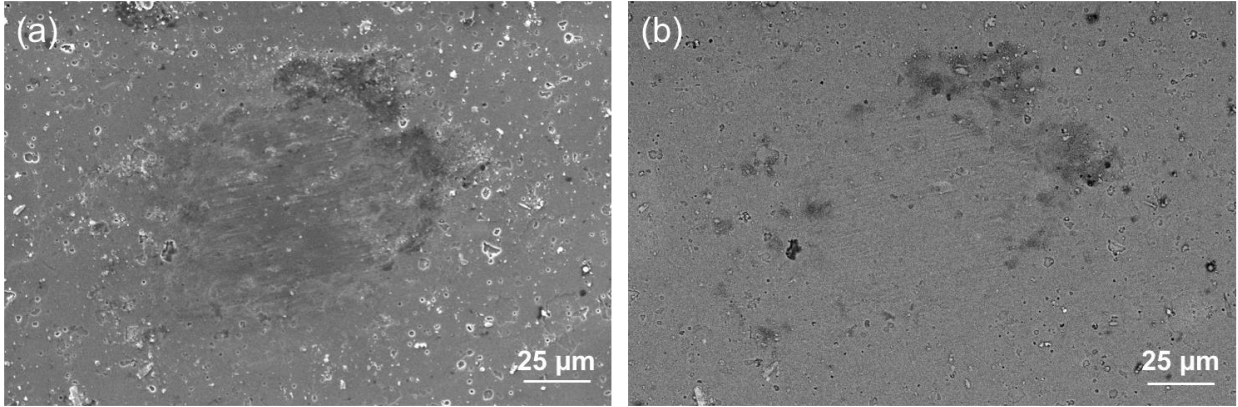
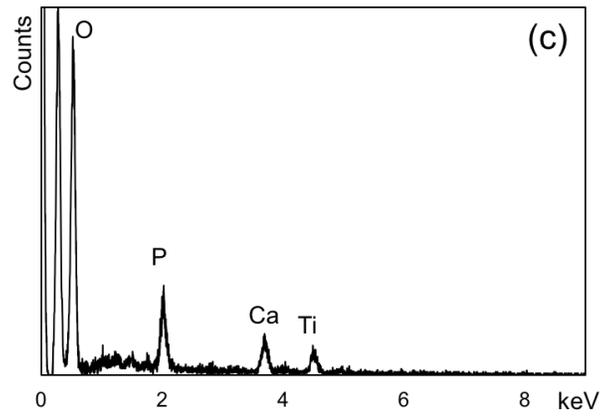

Figure 11. (a) SE and (b) BSE images of TiSiN coating after fretting test. (c) EDS spectrum of the fretting corrosion products.

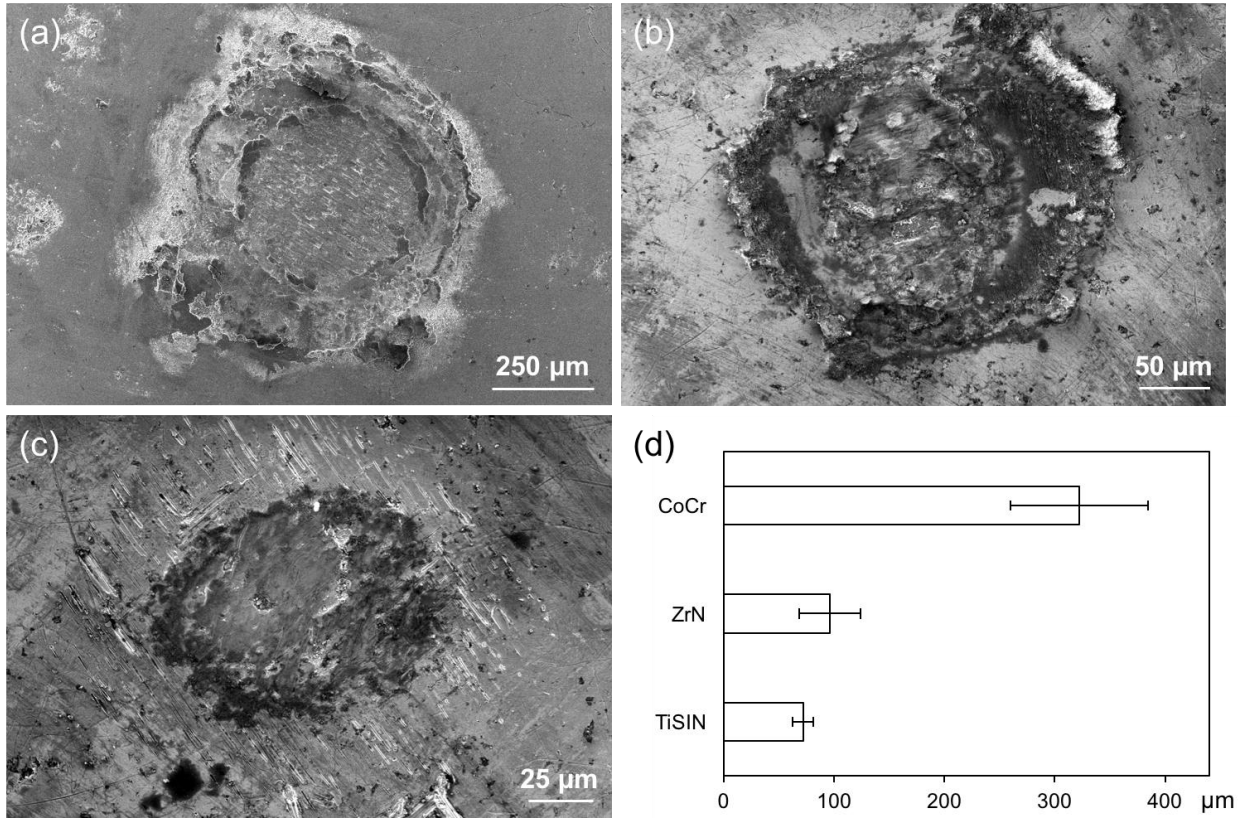

Figure 12. SEM images of Ti6Al4V spheres after fretting against (a) CoCrMo, (b) ZrN and (c) TiSiN. (d) Comparison of wear diameter on Ti6Al4V spheres from each group.

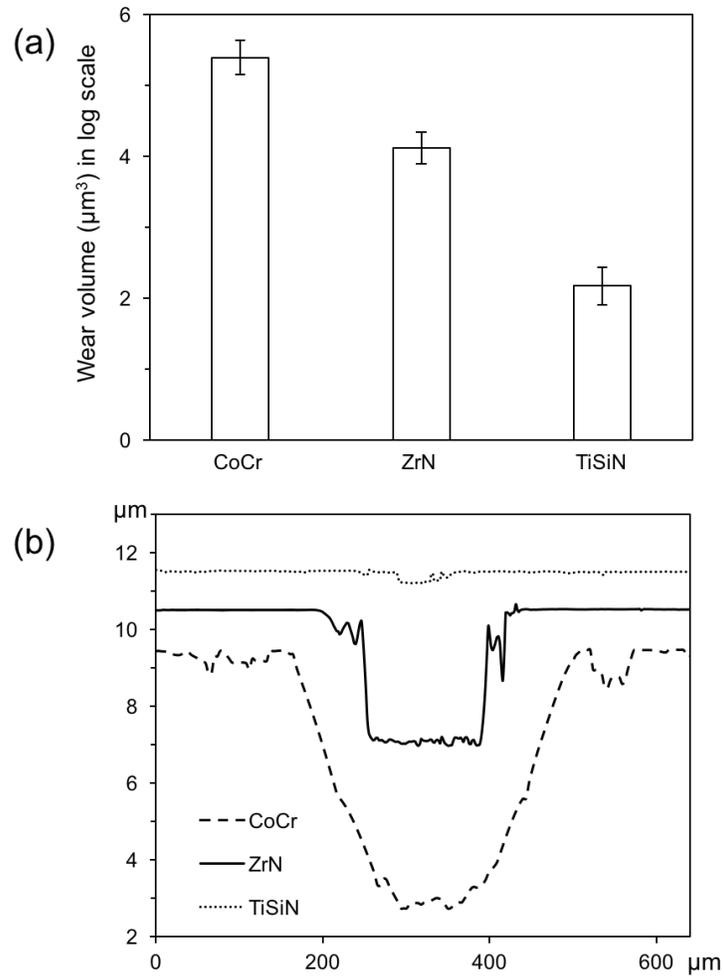

Figure 13. (a) Wear volume and (b) Wear track comparison of CoCr, ZrN and TiSiN after fretting against Ti6Al4V.

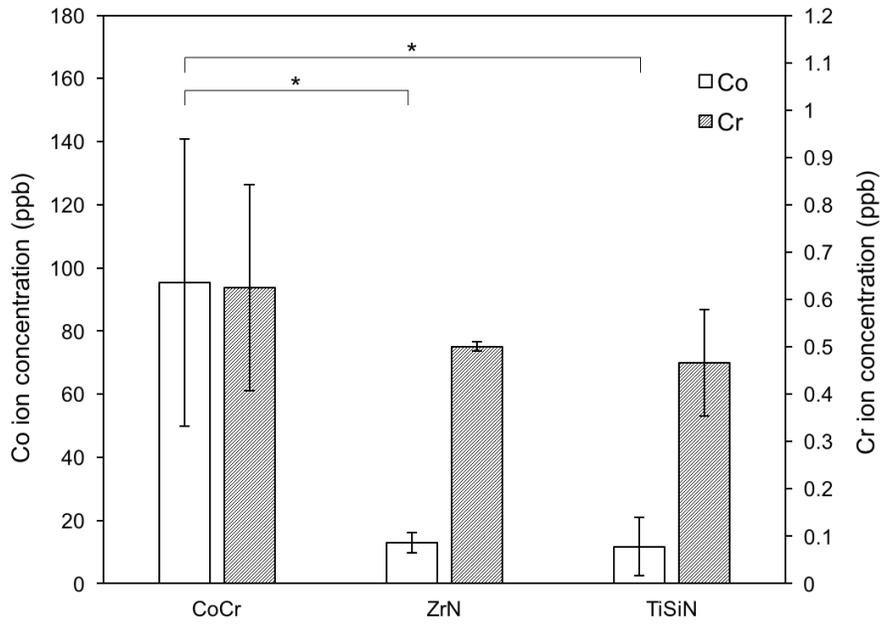

Figure 14. Concentrations of Co and Cr ion released during fretting test. Please note the different scales used for plotting Co and Cr ions. $p < 0.05$ is marked by *.

Table 1. Characteristics of ZrN, TiSiN coatings and the substrate CoCrMo

| Sample | Thickness (μm) | Surface roughness (nm) | Young's modulus (GPa) | Hardness (GPa) | Adhesion | Residual stress (GPa) |
|---|---|---|---|---|---|---|
| CoCrMo | N/A | 3.7 ± 0.3 | 180 ± 9 | 10.6 ± 1.2 | N/A | N/A |
| ZrN | 2.37 | 13.3 ± 2.1 | 409 ± 19 | 29.3 ± 2.0 | HF3 | -5.93 ± 0.10 |
| TiSiN | 1.89 | 40.6 ± 1.8 | 396 ± 29 | 41.6 ± 3.2 | HF3 | -8.00 ± 0.02 |

Table 2. Critical loads and failure mechanism of ZrN and TiSiN coatings

| Coating | Critical Loads (mN) | | Mechanism |
|---|---|---|---|
| ZrN | $Lc_1$ | 175.9 ± 8.8 | Tensile cracks |
| | $Lc_2$ | 262.3 ± 40.1 | Compressive delamination |
| TiSiN | $Lc_1$ | 207.8 ± 22.1 | Tensile cracks |
| | $Lc_2$ | 329.9 ± 20.6 | Chipping |

**Declaration of interests**

☒ The authors declare that they have no known competing financial interests or personal relationships that could have appeared to influence the work reported in this paper.

☐The authors declare the following financial interests/personal relationships which may be considered as potential competing interests:

## Author Statement

Chen-En Tsai: Investigation, Formal analysis, Writing- Original draft preparation
James Hung: Methodology-Coating processing, Formal analysis
Youxin Hu: Methodology-Fretting, Resources, Writing-Reviewing and Editing
Da-Yung Wang: Conceptualization, Methodology, Formal analysis, Supervision, Writing-Editing
Robert M. Pilliar: Conceptualization, Methodology, Writing-Reviewing and Editing
Rizhi Wang: Conceptualization, Methodology, Formal analysis, Supervision, Writing-Reviewing and Editing